\documentclass[aps,prl,showpacs,twocolumn,superscriptaddress]{revtex4}

\usepackage{amssymb, amsmath}
\usepackage{graphicx} 

\newcommand{\etal}{{\it et al.}}

\newcommand{\eqn}[1]{\label{eq:#1}}
\newcommand{\refeq}[1]{Eq.~(\ref{eq:#1})}
\newcommand{\fig}[1]{\label{fig:#1}}
\newcommand{\reffig}[1]{Fig.~\ref{fig:#1}}

\newcommand{\beq}{\begin{eqnarray}}
\newcommand{\eeq}{\end{eqnarray}}

\def \d {{\partial}}

\newcommand{\grad}{{\nabla}} 

\usepackage{ulem}
\usepackage[usenames]{color}

\begin{document}

\title{Swallowtail Band Structure of the Superfluid Fermi Gas in an Optical Lattice}        

\author{Gentaro Watanabe}
\affiliation{Asia Pacific Center for Theoretical Physics (APCTP),
Pohang, Gyeongbuk 790-784, Korea}
\affiliation{Department of Physics, POSTECH,
Pohang, Gyeongbuk 790-784, Korea}
\affiliation{Nishina Center, RIKEN, 2-1 Hirosawa, Wako, Saitama 351-0198, Japan}

\author{Sukjin Yoon}
\affiliation{Asia Pacific Center for Theoretical Physics (APCTP),
Pohang, Gyeongbuk 790-784, Korea}

\author{Franco Dalfovo}
\affiliation{INO-CNR BEC Center and Department of Physics, 
University of Trento, 38123 Povo, Italy}

\date{\today}      

\begin{abstract}
We investigate the energy band structure of the superfluid flow of
ultracold dilute Fermi gases in a one-dimensional optical lattice
along the BCS to BEC crossover within a mean-field approach. 
In each side of the crossover region, a loop structure (swallowtail) 
appears in the Bloch energy band of the superfluid above a critical 
value of the interaction strength. The width of the swallowtail is 
largest near unitarity. Across the critical value of the interaction 
strength, the profiles of density and pairing field change more 
drastically in the BCS side than in the BEC side. It is found that 
along with the appearance of the swallowtail, there exists a narrow 
band in the quasiparticle energy spectrum close to the chemical 
potential and the incompressibility of the Fermi gas consequently 
experiences a profound dip in the BCS side, unlike in the BEC side.
\end{abstract}

\pacs{03.75.Ss, 67.85.De, 67.85.Hj, 03.75.Lm}

\maketitle

Ultracold atoms in optical lattices attract much
interest because the controllability of both the lattice 
geometry and the interatomic interaction is such that they serve 
as testing beds for various models \cite{lattice}. For Bose-Einstein
condensates (BECs), it has been pointed out that the interaction 
can change the Bloch band structure drastically, causing 
the appearance of a loop structure called ``swallowtail'' in the energy 
dispersion \cite{wu_st,diakonov}. This is due to the competition between 
the external periodic potential and the nonlinear mean-field interaction: 
the former favors a sinusoidal band structure, while the latter tends to 
make the density smoother and the energy dispersion quadratic. When the 
nonlinearity wins, the effect of the external potential is screened and 
a swallowtail energy loop appears \cite{mueller}. This nonlinear 
effect requires the existence of an order parameter and, consequently, the 
emergence of swallowtails can be viewed as a peculiar manifestation of 
superfluidity in periodic potentials.

The problem of swallowtails can be even more important in Fermi 
superfluids due to the possible wide implications for various systems 
in condensed matter physics and nuclear physics. Indeed, extensive recent 
research activities are devoted to the simulation of solid states using 
cold Fermi gases and the behavior of cold fermions in optical lattices 
can lead to interesting analogies with superconductors and superconductor 
superlattices. In addition, our work may have implications 
also for superfluid neutrons in neutron stars, especially those in ``pasta'' 
phases (see, e.g., Ref.~\cite{qmd_formation} and references therein) in 
neutron star crusts \cite{vc_crossover}, where nuclei form a crystalline 
lattice in which superfluid neutrons can flow. However, unlike 
the Bose case \cite{wu_st,diakonov,mueller,machholm,BEC_Kronig,danshita}, 
little has been studied in this problem so far and a fundamental question of 
whether or not swallowtails exist along the crossover from the
Bardeen-Cooper-Schrieffer (BCS) to BEC states is still open.  In
this context, our work is aimed at showing the existence and the conditions 
for emergence of swallowtails in Fermi superfluids and presenting the 
unique features which make them different from those in bosons. 

We consider a two-component unpolarized dilute Fermi gas 
made of atoms of mass $m$ interacting with $s$-wave scattering 
length $a_s$ and subject to a one-dimensional (1D) optical lattice of 
the form
$V_{\rm ext}(\mathbf r)=sE_{\rm R} \sin^2q_{\rm B}z \equiv V_0 \sin^2q_{\rm B}z$.
Here, $V_0\equiv sE_{\rm R}$ is the lattice height, 
$s$ is the lattice intensity in dimensionless units, 
$E_{\rm R}=\hbar^2q_{\rm B}^2/2m$ is the recoil energy, 
$q_{\rm B}=\pi/d$ is the Bragg wave vector and $d$ is the lattice constant.
We compute the energy band structure of the system by solving the 
Bogoliubov-de Gennes (BdG) equations \cite{note_bdg}:
\beq
\left( \begin{array}{cc}
H(\mathbf r) & \Delta (\mathbf r) \\
\Delta^\ast(\mathbf r) & -H(\mathbf r) \end{array} \right)
\left( \begin{array}{c} u_i( \mathbf r) \\ v_i(\mathbf r)
\end{array} \right)
=\epsilon_i\left( \begin{array}{c} u_i(\mathbf r) \\
v_i(\mathbf r) \end{array} \right) \; ,
\eqn{BdG}
\eeq
where 
$H(\mathbf r) =-\hbar^2 \nabla^2/2m +V_{\rm ext}(\mathbf
r)-\mu$, 
$u_i(\mathbf r)$ and $v_i(\mathbf r)$ are quasiparticle
wavefunctions, and $\epsilon_i$ is the corresponding quasiparticle
energy.  The chemical potential $\mu$ is determined from the
constraint on the particle number $N= 2 \sum_i \int \left| v_i(\mathbf
r) \right|^2d{\bf r}$ and the pairing field $\Delta(\mathbf r)$ should 
satisfy a self-consistency condition $ \Delta(\mathbf r) =
-g \sum_i u_i(\mathbf r) v_i^*(\mathbf r)$,  where $g$ is the coupling 
constant for the $s$-wave contact interaction which 
needs to be renormalized. In the presence of a 
supercurrent with wavevector $Q=P/\hbar$ ($|P| \le P_{\rm edge} 
\equiv \hbar q_{\rm B}/2 $) moving in the $z$-direction \cite{note_current},
one can write the quasiparticle wavefunctions in the Bloch form as
$u_i(\mathbf r) =
\tilde{u}_i(z) e^{i Q z}e^{i\mathbf k \cdot \mathbf r }$ and
$v_i(\mathbf r) = \tilde{v}_i(z) e^{-i Q z}e^{i\mathbf k \cdot
\mathbf r }$
leading to the pairing field as 
$\Delta(\mathbf r)=e^{i 2Q z}\tilde{\Delta}(z)$.
Here $\tilde{\Delta}(z)$, 
$\tilde{u}_i(z)$, and $\tilde{v}_i(z)$ 
are complex functions with period $d$ and
the wave vector $k_z$ ($|k_z| \le q_{\rm B}$) lies in the 
first Brillouin zone. This Bloch decomposition transforms 
\refeq{BdG} into the following BdG equations for $\tilde{u}_i(z)$ 
and $\tilde{v}_i(z)$ :
\beq
\left( \begin{array}{cc}
\tilde{H}_{Q}(z) & \tilde{\Delta}(z) \\
\tilde{\Delta}^\ast(z) & -\tilde{H}_{-Q}(z) \end{array} \right)
\left( \begin{array}{c} \tilde{u}_i(z) \\ \tilde{v}_i(z)
\end{array} \right)
=\epsilon_i\left( \begin{array}{c} \tilde{u}_i(z) \\
\tilde{v}_i(z) \end{array} \right) \;,
\eqn{BdG2}
\eeq
where
\beq
  \tilde{H}_{Q}(z)\equiv \frac{\hbar^2}{2m} \left[ k^2_\perp
+\left(-i\partial_z+Q+k_z\right)^2 \right] +V_{\rm ext}(z) -\mu\, .
\nonumber\label{hq}
\eeq
Here, $k_\perp^2\equiv k_x^2 + k_y^2$ and
the label $i$ represents the wave vector $\mathbf k$ as well 
as the band index.

In the following, we mainly present the result for 
$s = V_0/E_{\rm R} = 0.1$ and $E_{\rm F}/E_{\rm R} =2.5$
as an example, where $E_{\rm F} = \hbar^2k_{\rm F}^2/(2m)$ 
and $k_{\rm F}=(3\pi^2 n_0)^{1/3}$ are the Fermi energy and 
momentum of a uniform free Fermi gas of density $n_0$.
These values fall in the range of parameters of feasible
experiments \cite{miller}.


\begin{figure}[!tb]
\centering
\resizebox{8.2cm}{!}
{\includegraphics{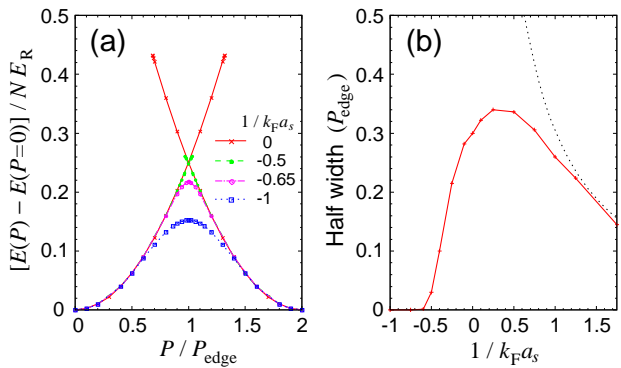}}
\caption{(Color online) (a) Energy $E$ per particle as a function of the
  quasimomentum $P$ for various values of $1/k_{\rm F}a_s$; 
  (b) half-width of the swallowtails along the BCS-BEC crossover. 
   These results are obtained for $s=0.1$ and $E_{\rm F}/E_{\rm R} 
  = 2.5$. The quasimomentum $P_{\rm edge}=\hbar q_{\rm B}/2$ fixes
  the edge of the first Brillouin zone. The dotted 
  line in (b) is the 
  half-width in a BEC obtained by solving the GP equation; it vanishes 
  at $1/k_{\rm F}a_s \simeq 10.6$.} \fig{swtail}
\end{figure}

We first compute the energy per particle in the lowest 
Bloch band as a function of the quasimomentum $P$ for various 
values of $1/k_{\rm F}a_s$. The results
in \reffig{swtail}(a) show that the swallowtails appear above a
critical value of $1/k_{\rm F}a_s$ where the interaction energy is strong 
enough to dominate the lattice potential. In \reffig{swtail}(b), the 
half-width of the swallowtails from the BCS to the BEC side is shown.
It reaches a maximum near unitarity ($1/k_{\rm F}a_s=0$).
In the far BCS and BEC limits, the width vanishes because the 
system is very weakly interacting and the band structure tends to 
be sinusoidal.
When approaching unitarity from either side, the interaction energy 
increases and can dominate over the periodic potential, which means that 
the system behaves more like a translationally invariant superfluid 
and the band structure follows a quadratic dispersion terminating
at a maximum $P$ larger than $P_{\rm edge}$. In the BEC side, we 
compare the results of our BdG calculations with those 
of the Gross-Pitaveskii (GP) equation for bosons of 
mass $m_b=2m$ interacting with scattering length $a_b=2a_s$ \cite{note_as}
in an optical lattice $2V_{\rm{ext}}(z)$: 
$-(\hbar^2/4m)\grad^2 \Phi(\mathbf{r}) + 2 V_{\rm{ext}}(z) \Phi(\mathbf{r}) 
+ (8\pi \hbar^2 a_s/2m) |\Phi(\mathbf{r})|^2 \Phi(\mathbf{r}) = 
\mu_B \Phi(\mathbf{r})$, where $\Phi(\mathbf{r})$ is a single 
macroscopic wavefunction describing the BEC. As it should be, the 
difference between the two curves becomes vanishingly small in 
the deep BEC regime. With our set of parameters ($s=0.1$ and 
$E_{\rm F}/E_{\rm R} =2.5$), the width of the swallowtail predicted by
the GP equation vanishes at $1/k_{\rm F}a_s\simeq 10.6$.


\begin{figure}[tb]
\resizebox{6cm}{!}
{\includegraphics{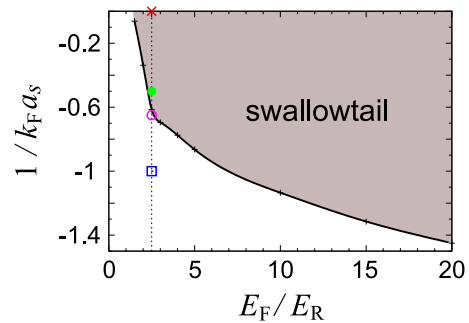}}
\caption{(Color online) 
The parameter region ($E_{\rm F}/E_{\rm R}$, $1/k_{\rm F} a_s$) 
where the swallowtails appear for $s=0.1$.
Symbols on the vertical line at $E_{\rm F}/E_{\rm R}=2.5$ 
correspond to the cases shown in \reffig{swtail}(a).
}
\fig{tail_cr}
\end{figure}

Whether the swallowtail appears or not depends on three parameters: the 
interaction parameter $1/k_{\rm F} a_s$, the lattice intensity $s$, and the 
ratio between the Fermi energy and the recoil energy $E_{\rm F}/E_{\rm R}$. 
In \reffig{tail_cr}, we fix $s=0.1$ and show the parameter region 
($E_{\rm F}/E_{\rm R}$, $1/k_{\rm F} a_s$) where we find swallowtails in the
BCS side of the crossover ($1/k_{\rm F} a_s<0$) \cite{note_landau}. 
We see that for weaker interaction (i.e., larger values of $1/k_{\rm F}|a_s|$) 
higher densities (larger values of $E_{\rm F}$) are required 
to create swallowtails, as expected.
The $s$-dependence of critical values of $1/k_{\rm F}a_s$ in the
BCS side is much weaker than the $1/s$ scaling behavior in the BEC side.
In the BCS side, as far as we have checked in $ 0.1 \le s \le 0.5$, 
the critical value of $1/k_{\rm F}a_s$ changes within
30\% at $E_{\rm F}/E_{\rm R} = 5$ and the change gets smaller with
increasing $E_{\rm F}/E_{\rm R}$.  This weak dependence of the
swallowtail region on $s$ in the BCS side is due to the Fermi
statistics: Provided $E_{\rm F}/V_0 = (E_{\rm F}/E_{\rm R})/s$ is
sufficiently larger than unity, the flow of the BCS
condensate formed from fermions near the Fermi surface is not very
sensitive to the presence of the lattice potential, while the
flow of the BEC formed from bosonic dimers all at
the bottom of the energy levels is more sensitive to it.


\begin{figure}[!tb]
\centering
\rotatebox{270}{
\resizebox{!}{8.5cm}
{\includegraphics{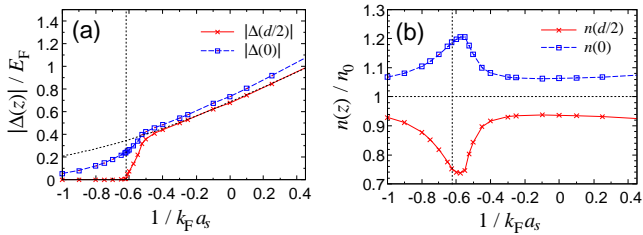}}}
\caption{(Color online) Profiles of (a) the pairing field $|\Delta(z)|$
  and (b) the density $n(z)$ along the change of $1/k_{\rm F}a_s$ for
  $P=P_{\rm{edge}}$ in the case of $s=0.1$ and $E_{\rm F}/E_{\rm R}
  =2.5$. The values of $|\Delta(z)|$ and $n(z)$ at the minimum ($z=0$,
  blue $\square$) and at the maximum ($z=\pm d/2$, red $\times$) of
  the lattice potential are shown.  The vertical dotted lines show the
  critical value of $1/k_{\rm F}a_s$ above which the swallowtail
  exists. The dotted curve in (a) shows $|\Delta|$ in the uniform system.
}
\fig{profiles}
\end{figure}

Both the pairing field and the density exhibit interesting 
features in the range of parameters where the swallowtails appear. 
This is particularly evident at the 
Brillouin zone boundary, $P=P_{\rm{edge}}$. In \reffig{profiles}, we show 
the magnitude of the pairing field $|\Delta(z)|$ and 
the density $n(z)$ calculated at the minimum ($z=0$) 
and at the maximum ($z=\pm d/2$) of the lattice potential. In general, $n(z)$ 
and $|\Delta(z)|$ take maximum (minimum) values where the external potential 
takes its minimum (maximum) values (for the full profiles, see Supplemental 
Material \cite{note_suppl}). The figure shows that $|\Delta(d/2)|$ 
remains zero in the BCS regime until the swallowtail appears at $1/k_{\rm F}a_s 
\approx -0.62$. Then it increases abruptly to values comparable to
$|\Delta(0)|$, which means that the pairing field becomes almost uniform 
at $P=P_{\rm{edge}}$ in the presence of swallowtails. As regards the density,
we find that the amplitude of the density variation, $n(0)-n(d/2)$, 
exhibits a pronounced maximum near the critical value of $1/k_{\rm F}a_s$.  
In contrast, in the BEC side,
the order parameter and the density are smooth monotonic 
functions of the interaction strength even in the region where the swallowtail 
appears. At $P=P_{\rm{edge}}$, the solution of the GP equation for bosonic
dimers gives the densities $n_b(0)= n_{b0}(1+V_0/2n_{b0} U_0)$ and 
$n_b(d/2)=n_{b0}(1-V_0/2n_{b0} U_0)$ with $V_0/2n_{b0}U_0 = (3\pi/4)
(sE_{\rm R}/E_{\rm F})(1/k_{\rm F}a_s)$, where $n_{b0}$ is the average 
density of bosons and 
$U_0 = 4\pi \hbar^2 a_b/m_b$ \cite{wu_st,bronski,note-nb}.  
Near the critical value of $1/k_{\rm F} a_s$, unlike the BCS side, the 
nonuniformity just decreases all the way even after the swallowtail appears. 
The local density at $z=d/2$ is zero until the swallowtail appears in the BEC
side while it is nonzero in the BCS side irrespective of the existence
of the swallowtail.
The qualitative behavior of $|\Delta(z)|$ around the critical
point of $1/k_{\rm F}a_s$ is similar to that of $n_b(z)$ because $n_b(\mathbf r)
= (m^2 a_s/8\pi)|\Delta(\mathbf r)|^2$ \cite{bdgtogp} in the BEC limit.

\begin{figure}[tb]
\centering
\resizebox{8.8cm}{!}
{\includegraphics{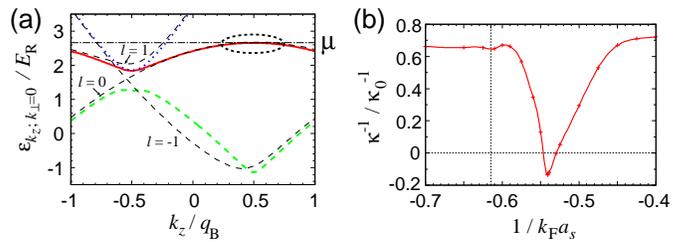}}
\caption{(Color online) (a) Lowest three Bloch bands of the quasiparticle 
energy spectrum at $k_{\perp}=0$ for $P=P_{\rm{edge}}$ 
and $1/k_{\rm F}a_s = -0.62$. 
Thin black dashed lines labeled by $l$'s
show the approximate energy bands obtained from
Eq.~(\ref{band_approx}) using $\mu \simeq 2.66 E_{\rm R}$
and $|\Delta|\simeq |\Delta(0)|\simeq 0.54 E_{\rm R}$.
(b) Incompressibility $\kappa^{-1}$ at $P=P_{\rm{edge}}$ around the 
critical value of $1/k_{\rm F}a_s \approx -0.62$ where the swallowtail starts 
to appear. The quantity $\kappa_0^{-1}$ is the incompressibility of the 
homogeneous free Fermi gas of the same average density. In both panels, we 
have used the values $s=0.1$ and $E_{\rm F}/E_{\rm R} =2.5$.
}
\fig{qp_spectrum}
\end{figure}


The quasiparticle energy spectrum plays an important role in 
determining the properties of the Fermi gas. Here we show that the emergence 
of swallowtails in the BCS side and for $E_{\rm F}/E_{\rm R}\agt 1$ is 
associated with peculiar structures of the quasiparticle energy spectrum
around the chemical potential. 
In the presence of a superflow moving in the $z$ direction with 
wavevector $Q$, the quasiparticle energies are given by the eigenvalues 
in \refeq{BdG2}. Since the potential is shallow ($s \ll 1$), some qualitative 
results can be obtained even ignoring $V_{\text{ext}}(z)$ except for its 
periodicity. With this assumption we obtain
\beq
\epsilon_{\mathbf k} \! \approx \! \frac{(k_z \! + \! 2q_{\rm B} l)Q}{m} \! + \! \sqrt{\left[ \! \frac{k_\perp^2 \! + \! (k_z \! + \! 2q_{\rm B} l)^2 \! + \! Q^2}{2m} \! - \! \mu \right]^2 \!\!\! + \! |\Delta|^2} \ ,
\label{band_approx}
\eeq 
with $l$ being integers for the band index. If $Q=0$, the $l=0$ band 
has the energy spectrum $\sqrt{[(k_\perp^2+ k_z^2)/2m - \mu]^2+ |\Delta|^2}$ 
which has a local maximum at $k_z=k_{\perp}=0$. When 
$Q=P_{\rm edge}/\hbar$, the spectrum is tilted and the local maximum moves 
to $k_z\simeq q_{\rm B}/2$ provided $|\Delta| \ll E_{\rm F}$ (and 
$E_{\rm F}/E_{\rm R}\agt 1$). 
In the absence of the swallowtail, the full BdG calculation indeed gives
a local maximum at $k_z=q_{\rm B}/2$ 
[see \reffig{qp_spectrum}(a), where
the green dashed line in the region $-1 < k_z/q_{\rm B}<-0.5$ and
the red solid line in the region $-0.5 < k_z/q_{\rm B} < 1$ 
correspond to the $l=0$ band] and the quasiparticle spectrum is
symmetric about this point, which 
reflects that the current is zero.
As $E_{\rm F}/E_{\rm R}$ increases, the band becomes flatter
as a function of $k_z$ and narrower in energy.

In \reffig{qp_spectrum}(a), we show the quasiparticle energy spectrum
at $k_{\perp}=0$ for $P=P_{\rm{edge}}$.  When the swallowtail is on the
edge of appearing, the top of the narrow band just
touches the chemical potential $\mu$ [see the dotted ellipse in
\reffig{qp_spectrum}(a)]. Suppose $1/k_{\rm F}a_s$ is slightly larger
than the critical value, so that the top of the band is slightly above
$\mu$. In this situation a small change of the quasimomentum $P$ causes a
change of $\mu$. In fact, when $P$ is increased
from $P=P_{\rm{edge}}$ to larger values, the band is tilted and the
top of the band moves upwards; the chemical potential $\mu$ should also
increase to compensate for the loss of states available. This implies
$\d\mu/ \d P> 0$. (See also Supplemental Material \cite{note_suppl}.) On the other hand, since the system is periodic, the
existence of a branch of stationary states with $\d\mu/ \d P> 0$ at
$P=P_{\rm{edge}}$ implies the existence of another symmetric branch with
$\d\mu/ \d P< 0$ at the same point, thus suggesting the occurrence of a
swallowtail structure.


A direct consequence of the existence of a narrow band in the
quasiparticle spectrum near the chemical potential is a strong reduction
of the incompressibility $\kappa^{-1} = n \d \mu(n)/\d n$ close to the
critical value of $1/k_{\rm F}a_s$ in the region where swallowtails exist
in the BCS side [see \reffig{qp_spectrum}(b)]. The dip
of $\kappa^{-1}$ occurs in the situation where the top of the narrow band
is just above $\mu$ for $P=P_{\rm edge}$
($1/k_{\rm F}a_s$ is slightly above the critical value). 
An increase of the density $n$ has
little effect on $\mu$ in this case, because the density of states is
large in this range of energy and the new particles can easily adjust
themselves near the top of the band by a small increase of $\mu$.
This implies that $\d \mu(n)/\d n$ is small and the incompressibility has a
pronounced dip \cite{note_incompress}. It is worth noting that in
the BEC side the appearance of the swallowtail is not associated with any
significant change of incompressibility. In fact, the exact solution of
the GP equation gives $\kappa^{-1} = n_{b0}U_0$
near the critical conditions for the occurrence of swallowtails,
being a smooth and monotonic function of the interaction strength.

Swallowtails may produce observable effects in the behavior of Bloch 
oscillations \cite{landau-zener,diakonov}. Since Bloch oscillations 
have various important applications, such as precision measurements of 
forces \cite{roati} and controlling the motion of a wave packet
\cite{alberti,haller}, a better understanding of swallowtails in bosonic and 
fermionic gases is certainly useful in these contexts. One may 
also exploit the tunability of the interaction and the peculiar 
dynamics of the superfluid in the lattice for different applications.  
For instance, by periodically sweeping a magnetic field across 
the critical region for the appearance of swallowtails, one can 
produce a time modulation of the shape of the lowest energy band 
between a sinusoidal form and a quadratic-like form.  Since the absolute 
value of the group velocity $\partial_P[E(P)/N]$ of the latter is 
always larger than that of the former [see \reffig{swtail}(a)], one 
could realize a directed motion of the gas by synchronizing the period 
of this modulation with the period of the Bloch oscillations
(see also Supplemental Material \cite{note_suppl}).  
This may be experimentally more accessible in the BCS side than in the 
BEC side because the critical value of $1/k_{\rm F}|a_s|$ is of order 
$1$ for a wide range of $s$ and $E_{\rm F}/E_{\rm R}$. 
This new method would complement other proposals for realizing
directed motion of atomic wave packets in 1D optical lattices
\cite{directmotion,haller}.

In summary, we have predicted the existence of swallowtails 
in the energy band of superfluid fermions in a lattice and have
pointed out some key features which make these swallowtails 
different from those in a BEC. The results are obtained within a 
range of parameters compatible with current experiments \cite{miller}.
We hope our predictions stimulate experiments aimed to
observe swallowtails with Fermi gases.

\begin{acknowledgments}
We acknowledge C. J. Pethick, Y. Shin, and T. Takimoto for helpful discussions.
This work was supported by the Max Planck Society, 
the Korea Ministry of Education, Science and Technology, 
Gyeongsangbuk-Do, Pohang City, for the support of 
the JRG at APCTP, and ERC through the QGBE grant.
Calculations were performed on RICC in RIKEN and Wiglaf 
at the University of Trento. 
\end{acknowledgments}

\clearpage

\appendix

\begin{widetext}
\section{Supplemental Material for Swallowtail Band Structure of the Superfluid Fermi Gas in an Optical Lattice}
\begin{center}
Gentaro Watanabe, Sukjin Yoon, and Franco Dalfovo
\end{center}
\end{widetext}






\section{Full profiles of the pairing field $|\Delta(z)|$ and the density $n(z)$}

In Fig.~3 of the paper, the values of $|\Delta(z)|$ and $n(z)$ at the
minimum and at the maximum of the lattice potential are shown. The
full profiles of $|\Delta(z)|$ and $n(z)$ along the lattice vector
($z$-direction) are given in the following \reffig{profiles_supp}.
By increasing the interaction parameter $1/k_{\rm F}a_s$, we find that
the order parameter $|\Delta|$ at the maximum ($z=\pm d/2$) of the lattice
potential exhibits a transition from zero to nonzero values at the
critical value of $1/k_{\rm F}a_s$ at which the swallowtail appears
[see also Fig.\ 3(a) of the paper].
Note that here we plot the absolute value of $\Delta$; the order
parameter $\Delta$ behaves smoothly and changes sign across zero.

\begin{figure}[!htb]
\centering
\resizebox{7.5cm}{!}
{\includegraphics{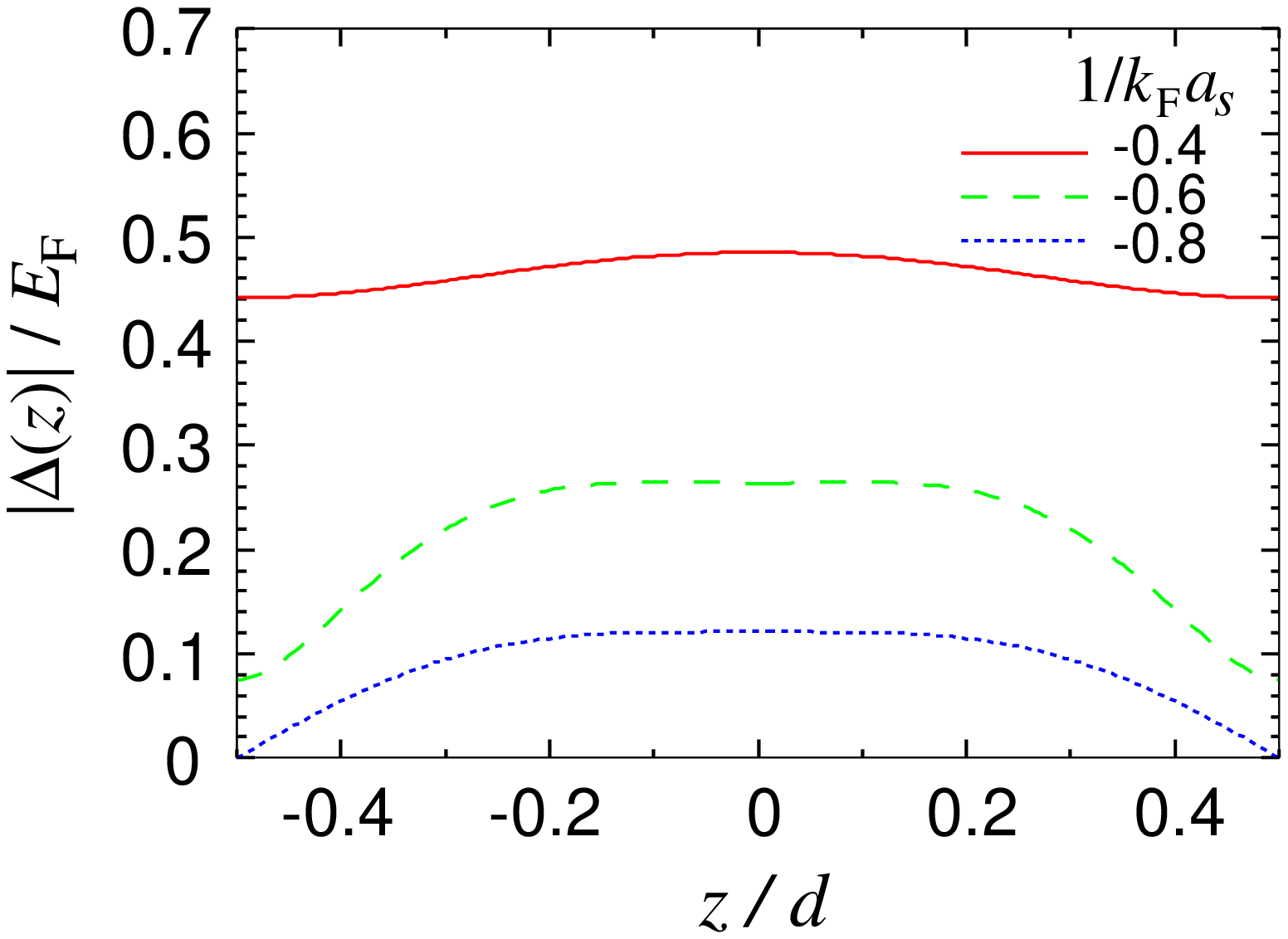}}
\resizebox{7.5cm}{!}
{\includegraphics{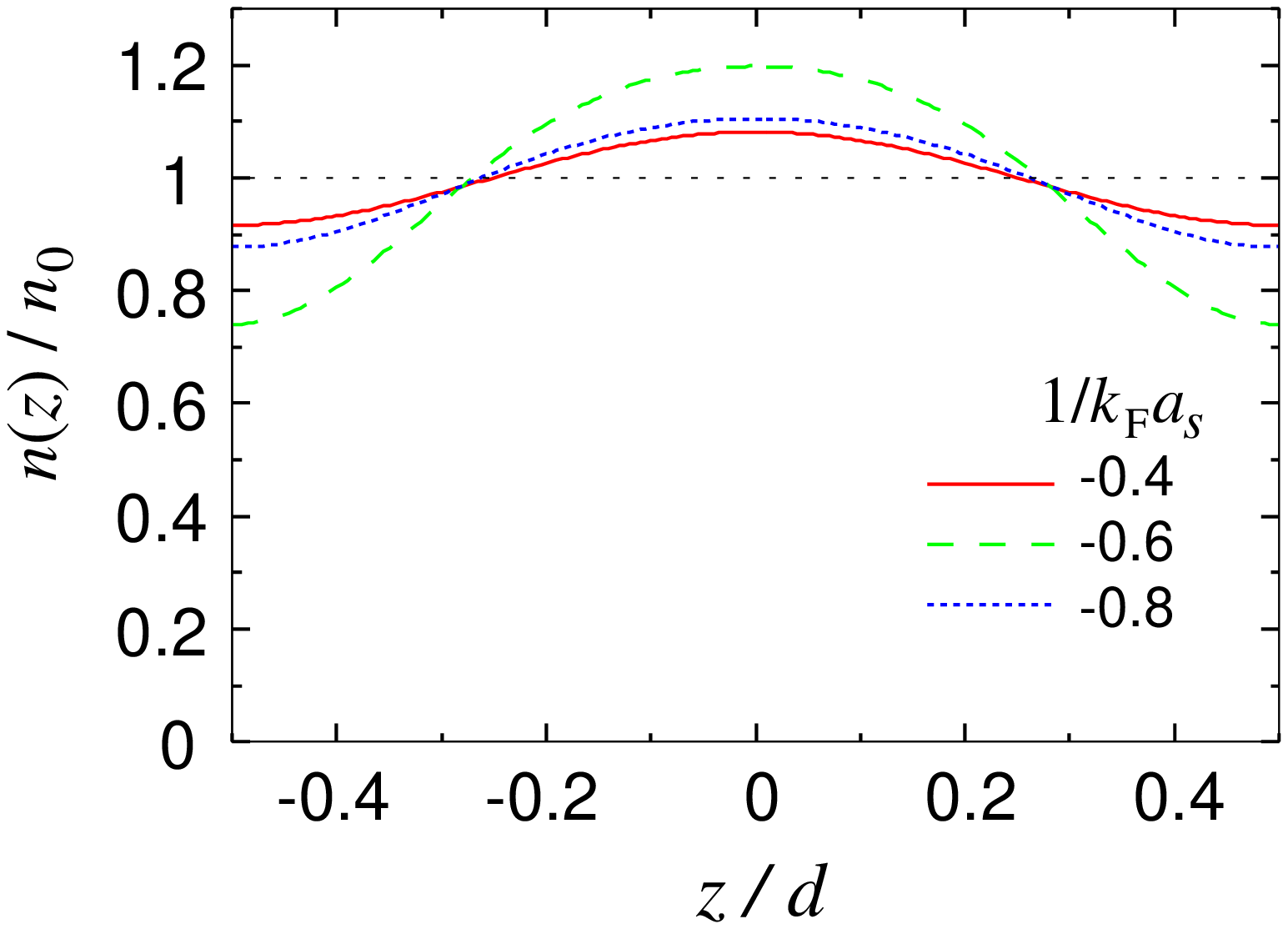}}
\caption{Profiles of the pairing field $|\Delta(z)|$ and the density $n(z)$ at
  $1/k_Fa_s= -0.8$ (blue dotted), $-0.6$ (green dashed), and $-0.4$
  (red solid) for $P=P_{\rm{edge}}$ in the case of $s=0.1$ and 
  $E_{\rm F}/E_{\rm R} =2.5$.  The swallowtail starts to appear
  at a critical value of $1/k_{\rm F}a_s \approx -0.62$.
}
\fig{profiles_supp}
\end{figure}

The blue dotted, green dashed, and red solid lines correspond to the
cases where the interparticle interaction strengths are short, just
enough, and sufficiently large for developing the swallowtails,
respectively.
\newline

\section{Behavior of the Bloch Band near the value of the chemical potential with the change of the quasimomentum}

In Fig. 4(a) of the paper, the lowest three Bloch bands of the
quasiparticle energy spectrum for $P=P_{\rm{edge}}$ at $k_{\perp}=0$
and $1/k_{\rm F}a_s = -0.62$ are given. To visualize the behavior of
the second band (the red solid line in that figure) near the value of the
chemical potential with the change of the quasimomentum $P$, we show
the cases of $P/P_{\rm{edge}}= 0$ (black dotted), $0.5$ (green
dashed), and $1$ (red solid) in the following \reffig{qp_spect_suppl}.
Notice that the sharp minima in the curves for $P/P_{\rm edge}=0$ and $-0.5$
are due to avoided crossings with other bands.

\begin{figure}[!htb]
\centering
\resizebox{7.5cm}{!}
{\includegraphics{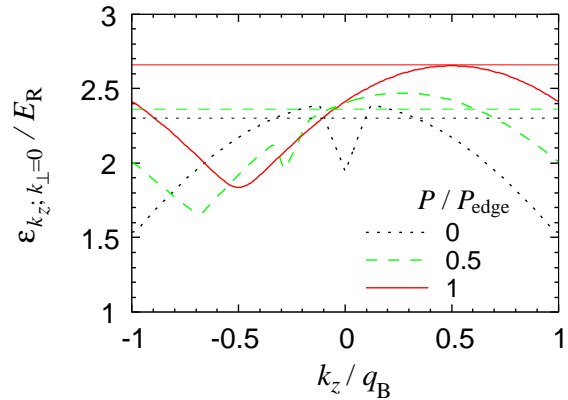}}
\caption{Bloch band of the quasiparticle energy spectrum around the
  chemical potential for $P/P_{\rm{edge}}=0$ (black dotted), $0.5$
  (green dashed), and $1$ (red solid), respectively, at $k_{\perp}=0$
  and $1/k_{\rm F}a_s = -0.62$ in the case of $s=0.1$ and 
  $E_{\rm F}/E_{\rm R} =2.5$. The each horizontal line denotes the value of
  the chemical potential for the corresponding value of $P$.
}
\fig{qp_spect_suppl}
\end{figure}

\section{Application of the swallowtail band structure to a Directed motion}

\begin{figure}[!htb]
\centering
\resizebox{7.cm}{!}
{\includegraphics{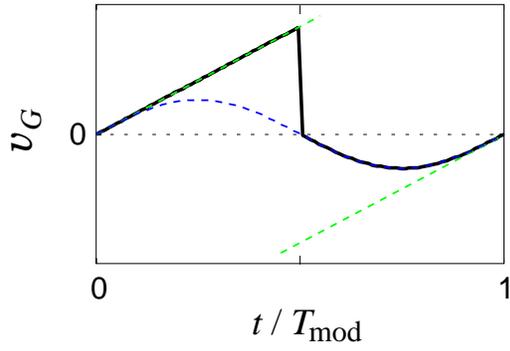}}
\caption{Schematic diagram of the group velocity $v_G$ during one
  period ($T_{\rm mod}$) of modulation of the magnetic field
  (accordingly, $1/k_{\rm F}a_s$). The green (blue) dashed line shows
  the group velocity along the time when $1/k_{\rm F}a_s$ is set to
  the value for a quadratic (sinusoidal) band structure as depicted in
  Fig.\ 1(a) of the paper. When $1/k_{\rm F}a_s$ is switched between
  the two values in a proper way, the group velocity of a wave packet
  will follow the black solid line.  In real situation, the black line
  around $t=T_{\rm mod}/2$ is a smooth curve rather than a sharp
  drop due to a continuous variation of $1/k_{\rm F}a_s$ connecting
  the two values.} \fig{directed_motion_suppl}
\end{figure}

In \reffig{directed_motion_suppl}, we show a schematic diagram for our
proposal of the directed motion of an atomic wave packet. The basic
idea is making use of the difference in the group velocity $v_G$
between the quadratic-like band with a swallowtail and the
sinusoidal-like band.  By modulating $1/k_{\rm F}a_s$ in such a way
that the dispersion is quadratic when $v_G$ is positive and sinusoidal
when $v_G$ is negative, the net displacement over one period of
modulation of $1/k_{\rm F}a_s$ is positive and a directed motion can
be produced.


\begin{thebibliography}{99}
\bibitem{note_bdg} Neither the hydrodynamic theory nor the tight-binding model 
is appropriate since they cannot describe the
swallowtails correctly along the BCS-BEC crossover: the former always
yields the dispersion of the quadratic form without termination and
the latter gives the sinusoidal band.
\bibitem{note_current} Nonzero values of $P$ can be obtained,
for instance, by orienting the periodic potential in the direction of 
the gravity field or by imposing a suitable time-dependent phase to 
the optical lattice.
\bibitem{note-nb} This result is exact only when the swallowtails exist, 
otherwise it is approximate, but qualitatively correct.
\bibitem{note_incompress} With the parameters used in 
\reffig{qp_spectrum}(b), the incompressibility takes negative values
in a small region around $1/k_{\rm F}a_s =-0.55$, which means that the system 
might be dynamically unstable against long-wavelength perturbations.
By choosing appropriate parameters, this region disappears.
\end{thebibliography}
\end{document}